\documentclass[onecollarge,natbib]{svjour2}
\bibpunct{[}{]}{,}{n}{}{,} 
\smartqed  
\usepackage{graphicx}
\usepackage{amsmath,amssymb}

\newcommand{\ubar}{\text{$\overline{{\rm u}}$}}
\newcommand{\dbar}{\text{$\overline{{\rm d}}$}}
\newcommand{\cbar}{\text{$\overline{{\rm c}}$}}
\newcommand{\Dbar}{\text{$\overline{{D}}$}}
\newcommand{\Jpsi}{{J\!/\!\psi}}

\newcommand{\X}{\text{X(3872)}}
\newcommand{\DDbarz}{$D^0\Dbar{}^{*0}$}
\newcommand{\DDbarpm}{$D^+D^{*-}$}
\newcommand{\ccbar}{{\rm c}\cbar}

\newcommand{\bra}{\langle}
\newcommand{\ket}{\rangle}
\newcommand{\rmd}{{\rm d}}
\newcommand{\rmi}{{\rm i}}

\newcommand\GinP{G^{(P)}}

\renewcommand{\Im}{{\rm Im}~}
\newcommand{\xbld}[1]{\mbox{\boldmath $#1$}}

\newcommand{\vecp}{\xbld{p}}
\newcommand{\kf}{k_{\!f}}

\newcommand{\Belle}{\text{Belle}}
\newcommand{\BABAR}{\text{\it{\hspace{-.1em}B\hspace{-.1em}A\hspace{-.1em}B\hspace{-.1em}A\hspace{-.1em}R\hspace{.1em}}}}

\journalname{Few-Body Systems (APFB2011)}
\begin{document}

\title{\boldmath
Charmonium and meson-molecule hybrid tetraquarks
}
\subtitle{Vector meson width and the isospin breaking in the \X\ decay}


\author{
Sachiko Takeuchi
\and
Kiyotaka Shimizu
\and
Makoto Takizawa
}


\institute{S.\ Takeuchi \at
Japan College of Social Work, Kiyose, Tokyo 204-8555, Japan \\
              \email{s.takeuchi@jcsw.ac.jp}           
           \and
K.\ Shimizu \at
Department of Physics, Sophia University, Chiyoda-ku, Tokyo 102-8554, Japan
\and
M.\ Takizawa \at
Showa Pharmaceutical University, Machida, Tokyo 194-8543, Japan
}

\date{Received: date / Accepted: date}

\maketitle

\begin{abstract}
In the \X\ decay, both of the $\Jpsi\pi\pi$ and $\Jpsi\pi\pi\pi$  branching fractions
 are observed experimentally, 
and their sizes are comparable to each other.
In order to clarify the mechanism to cause such a large isospin violation,
we investigate  \X\ employing a model of coupled-channel two-meson 
scattering with a \ccbar\ core.
The two-meson states consist of 
\DDbarz, \DDbarpm,
$\Jpsi\rho$, and 
$\Jpsi\omega$.
The effects of the $\rho$ and $\omega$  meson width are also taken into account.

We calculate the transfer strength from
the \ccbar\ core to 
the final two-meson states.
It is found 
that
very narrow $\Jpsi\rho$ and  
$\Jpsi\omega$ peaks appear very close to the \DDbarz\ threshold
for a wide range of variation in the parameter sets.
The size of the $\Jpsi\rho$ peak is almost the same as that of 
$\Jpsi\omega$, 
which is consistent with the experiments.
The large width of the $\rho$ meson 
 makes the originally small isospin violation by about five times larger.
\end{abstract}

\section{Introduction}
\label{intro}
\abovedisplayskip=4.5pt
\belowdisplayskip=4.5pt
\allowdisplaybreaks[4]

\X\ has been found first by \Belle\ \cite{Choi:2003ue}
and then confirmed by
various experiments \cite{pdg,Brambilla:2010cs}.
The mass of \X\ is found to be 3871.57$\pm$0.25 MeV,
which is almost on the \DDbarz\ threshold, 3871.73 MeV.
The width is less than 2.3 MeV, 
which is very narrow for such a highly excited resonance,
and $J^{PC}$=$1^{++}$
or $2^{-+}$  \cite{pdg}. 
One of the peculiar features of  \X\ is that 
the isospin mixing occurs on a large scale.
The decay branching fraction of \X\ to three pions
is comparable to that of two pions \cite{Abe:2005ix,delAmo}:
\begin{eqnarray}
{Br(X\rightarrow \pi^+\pi^-\pi^0 \Jpsi)
\over 
Br(X\rightarrow \pi^+\pi^-\Jpsi)}
&=&1.0 \pm 0.4 \pm 0.3 ~~~(\Belle)
~~~~=~
0.8\pm 0.3 ~~~(\BABAR).
\end{eqnarray}

It has been pointed out in many works that \X\ is not a simple \ccbar\ state 
\cite{pdg,Brambilla:2010cs}. 
The observed \X\ mass is by about 78 MeV lower 
than the $J^{PC}$=$1^{++}$ \ccbar\ state predicted by the quark model,
which has successfully explained 
the \ccbar\ mass spectrum below the $D\Dbar$ threshold.
As seen in Table \ref{tbl:meson-mass}, 
there are four thresholds which are very close to the \X\ mass.
It is natural to consider that \X\ has large components of these two-meson states.
The spectrum of the final pions,
the radiative decay modes,
the production rate, however,
suggest that \X\ has a \ccbar\ core.
We argue that \X\ is a hybrid state of the \ccbar\ core and the two-meson molecule
with $J^{PC}$=$1^{++}$.

%
\begin{table}[t]
\caption{Mass and width of the mesons and relevant thresholds (in MeV). Data are taken from Ref.\cite{pdg}.}
\centering
\label{tbl:meson-mass}
\tabcolsep=0.7mm
\begin{tabular}{ccccccc@{~~~~~}ccccccc}\hline\noalign{\smallskip}
$D^0$ & $D^{*0}$ & $\Jpsi$ & $\rho~(\Gamma_\rho)$ & $\omega~(\Gamma_\omega)$ & $D^+$ & $D^{*+}$ 
&$D^0$-$\Dbar^{*0}$ & $\Jpsi$-$\rho$ & $\Jpsi$-$\omega$ & $D^+$-$D^{*+}$
\\[3pt]
\tableheadseprule\noalign{\smallskip}
1864.80 & 2006.93 & 3096.92 & 775.49(149.1) & 782.65(8.49)& 1869.57 & 2010.22
&3871.73 & 3872.41 & 3879.57 & 3879.79
\\ 
\noalign{\smallskip}\hline
\end{tabular}\vspace*{-1mm}
\end{table}%

\begin{table}[t]
\caption{Model parameters for the interaction.}
\label{tbl:param}
\begin{center}
\def\TenZ\phantom{.0}
\begin{tabular}{rcccccc}\hline\noalign{\smallskip}
Parameter set &$V^{(0)}_{PP}$(MeV$\cdot$fm$^3$)& $V^{(0)}_{QP}$(MeV$\cdot$fm$^{3/2}$) &$a$(fm) &$c$& $E_0^{(Q)}$ (MeV)\\[3pt]
\tableheadseprule\noalign{\smallskip}
A~~ & 3\phantom{.00} &10 & 0.4 & ~~1\phantom{.0} & 3950\\ 
B~~ & 3.86 & 10 & 0.4&~~0.5&3950\\
\noalign{\smallskip}\hline
\end{tabular}\vspace*{0mm}
\end{center}
\end{table}

\section{Model Hamiltonian, the Lippmann-Schwinger equation, the transfer strength}
\label{sec:1}

The two-meson channels we consider are 
\DDbarz,
\DDbarpm,
$\Jpsi\rho$ , and 
$\Jpsi\omega$.
We introduce the orbitally excited \ccbar\ core, which is treated as the
bound state embedded in
the continuum (BSEC) \cite{Takeuchi:2008wc}.
The two-meson states and the 
\ccbar\ core are denoted by $P$-space and  $Q$-space, respectively.

The model Hamiltonian 
 is:
\begin{eqnarray}
H&=&\left(\begin{array}{cc}
H_{PP} & V_{PQ}\\
V_{QP}&E^{(Q)}_0
\end{array}
\right)=H_0+V
~~~~\text{with}~~~
H_0=
\left(\begin{array}{cc}
H^{(P)}_0 & 0\\
0&E^{(Q)}_0
\end{array}
\right)
~~~~\text{and}~~~
V=
\left(\begin{array}{cc}
V_{PP} & V_{PQ}\\
V_{QP}&0
\end{array}
\right)
\end{eqnarray}
where $H_{PP}$ is the Hamiltonian for the two-meson systems,
$V_{PQ}$ and $V_{QP}$ are the transfer potentials 
and $E^{(Q)}_0$ is the BSEC mass before the coupling to the scattering states is switched on.
We take its value from the prediction by the quark model.

Since the concerning particles are rather heavy and 
we consider the energy region very close to the threshold, 
we use a separable potential
for the two-meson systems with the nonrelativistic treatment:
\begin{eqnarray}
 H^{(P)}_0&=&\sum_i  \left(M_i+m_i+ {k_i^2\over 2 \mu_i}
\right)\\[-0.5em]
V_{PP;ij}(\vecp,\vecp')&=&
V^{(0)}_{PP}~\lambda_{ij}~
g(p)g(p')Y_{00}(\Omega_p)Y^*_{00}(\Omega_p')~~~~~\text{with}~~~~~
g(p)=\exp [-\frac{{a^2}p^2}{4}]
\label{eq:g}
\end{eqnarray}
where $M_i$ and $m_i$ are the masses of the two mesons of the $i$-th channel, 
 $ \mu_i$ is their reduced mass, and
 $k_i$ is their relative momentum.
As for the range parameter, $a$, 
of the gaussian separable potential between the two mesons,
$V_{PP}$,
 we use a common value
for all the channels for simplicity.
$V^{(0)}_{PP}$ is the strength of the interaction, and 
the factor $\lambda_{ij}$ describes the channel dependence.

The transfer potentials between \ccbar\ and the two-meson states, 
$V_{PQ}$ are taken as:
\begin{eqnarray}
\bra Q|V_{QP;i}|\vecp\ket  
&=&
V^{(0)}_{tr}~\tilde{\lambda_i}~{g}(p)Y^*_{00}(\Omega_p)
\end{eqnarray}
where
$V^{(0)}_{tr}$ is the strength of the transfer potential, and 
the factor $\lambda_{i}$ stands for the channel dependence.

The potential $V$  is chosen not to violate the isospin symmetry.
In addition, 
the direct coupling between \DDbarz(c\ubar-u\cbar) and \DDbarpm(c\dbar-d\cbar) 
is assumed not to occur because it is forbidden by the OZI rule.
So, the $V_{PP}$ for the  $D\overline{D}{}^*$ isospin 1 system becomes the same as that of the isospin 0 system.
We also assume that the \ccbar\ core 
does not directly couple to $\Jpsi\rho$ nor $\Jpsi\omega$
because the transfer is forbidden again by the OZI rule.
Moreover, the interaction between $\Jpsi$ and the vector meson is assumed to vanish,
as it vanishes at the heavy quark mass limit.
Thus, 
the channel dependence of the interaction is summarized as \vspace*{-2mm}
\begin{eqnarray}
\{\lambda_{ij}\} &=&  \left( \begin{array}{rrrr}
-1&  0&  c& ~~c\\
 0& -1& -c& c\\
 c& -c&  0& 0\\  
 c&  c&  0& 0
 \end{array}\right)
~~~~\text{and}~~~
\{\tilde{\lambda}_i\} 
=
 \left( \begin{array}{rrrr}
 1&  1&  0& 0
 \end{array}\right)
\end{eqnarray}
for the \DDbarz, \DDbarpm, $\Jpsi\rho$, and $\Jpsi\omega$ channels, respectively.
The strengths of the potentials have been investigated but not well understood yet.
Here  we choose the values of $V^{(0)}_{PP}$ and $V^{(0)}_{QP}$ 
so that the \X\ becomes a peak at the observed energy.
The potential is considered to be short-ranged, so we use 
a typical size, 0.4 fm,  for the range  parameter.
As for the parameter $c$, we choose two different values 
are shown in Table \ref{tbl:param}.

We first solve the  system without introducing the width of the vector mesons:
\begin{eqnarray}
T&=&V+VG_0T
~~~~\text{with}~~~
T=
\left(\begin{array}{cc}
T_{PP} & T_{PQ}\\
T_{QP}&T_{QQ}
\end{array}
\right)
~~~~\text{and}~~~
G_0=
\frac{1}{E-H_0+{\rm i}\varepsilon} .
\end{eqnarray}

The `full' propagator solved within the $P$-space, $\GinP$, 
and the full propagator of $Q$, $G_Q$, 
can be obtained as
\begin{eqnarray}
\GinP &=&
 \left(E-H^{(P)}_0-V_{PP}+{\rm i}\varepsilon\right)^{-1}
 ~~~~\text{and}~~~~
G_Q
=
\left(E-E_0^{(Q)}-V_{QP}\GinP V_{PQ}\right)^{-1}
\end{eqnarray}

It is considered that  
\X\ is produced via the
\ccbar\ state in the B meson decay.
Thus,  the observed mass spectrum is proportional to
 the transfer strength from the \ccbar\ core to the final meson states:
\begin{eqnarray}
{1\over c_K }{\rmd W\over \rmd E} &=& 
\sum_f \mu_{f}\kf |\bra f;\kf |  T_{PQ}G_0 |\text{\ccbar}\ket |^2
=
-{1\over \pi}\Im\bra \text{\ccbar} | (T_{PQ}G_0)^\dag  \GinP_0 T_{PQ} G_0 |\text{\ccbar} \ket
\label{eq:17}
\end{eqnarray}
where $c_K$ is a factor which comes from the kaon phase space,
$E$ is the energy of \DDbarz\ when the center of mass of the two mesons is at rest.
The strength for the open channel $f$ can be rewritten as
\begin{eqnarray}
{1\over c_K }{\rmd W(\text{\ccbar}\rightarrow f)\over \rmd E}
&=&
\mu_{f}\kf \left| \bra f;\kf |(1+V_{PP}\GinP)V_{PQ}G_Q |\text{\ccbar} \ket \right|^2~.
\end{eqnarray}


Next we introduce the effects of the $\rho$ and $\omega$ meson width.
We assume that the decay of the vector mesons  occurs only at the final two-meson states.
Namely, we modify the free propagator within the 
$P$-space $\GinP_0$ in  eq.\ (\ref{eq:17}) as\vspace*{-1ex}
\begin{eqnarray}
\GinP_0\rightarrow \tilde{G}^{(P)}_0&=& 
\left\{E-(M_i+m_i+\frac{k^2}{2\mu_i})+{\rmi \over 2} \Gamma_V(s(k))\right\}^{-1}
\end{eqnarray}
The width of the vector mesons,  $\Gamma_V(s)$, depends on the energy of the pion relative motion, $s$, which depends on $k$.
The parameters in $\Gamma_V$ are taken so that it produces
the observed  $\rho$ or $\omega$ width.

Thus we have the strength for the open channel $f$  as
\begin{eqnarray}
{1\over c_K }{\rmd W(\text{\ccbar}\rightarrow f)\over \rmd E}
&=&
{2\over \pi} \mu_{f} 
\int {k^2 \rmd k~\mu_f\Gamma_V(s(k))
\over
(\kf^2-k^2)^2 +(\mu_f\Gamma_V(s(k)))^2}
 \left| \bra f;k|(1+V_{PP}\GinP)V_{PQ}G_Q |\text{\ccbar} \ket \right|^2~.
\label{eq:eq32}
\end{eqnarray}
The detailed calculation is given in ref.\ \cite{Takeuchi:2008wc}.

\begin{figure}
\centering
  \includegraphics[scale=0.4]{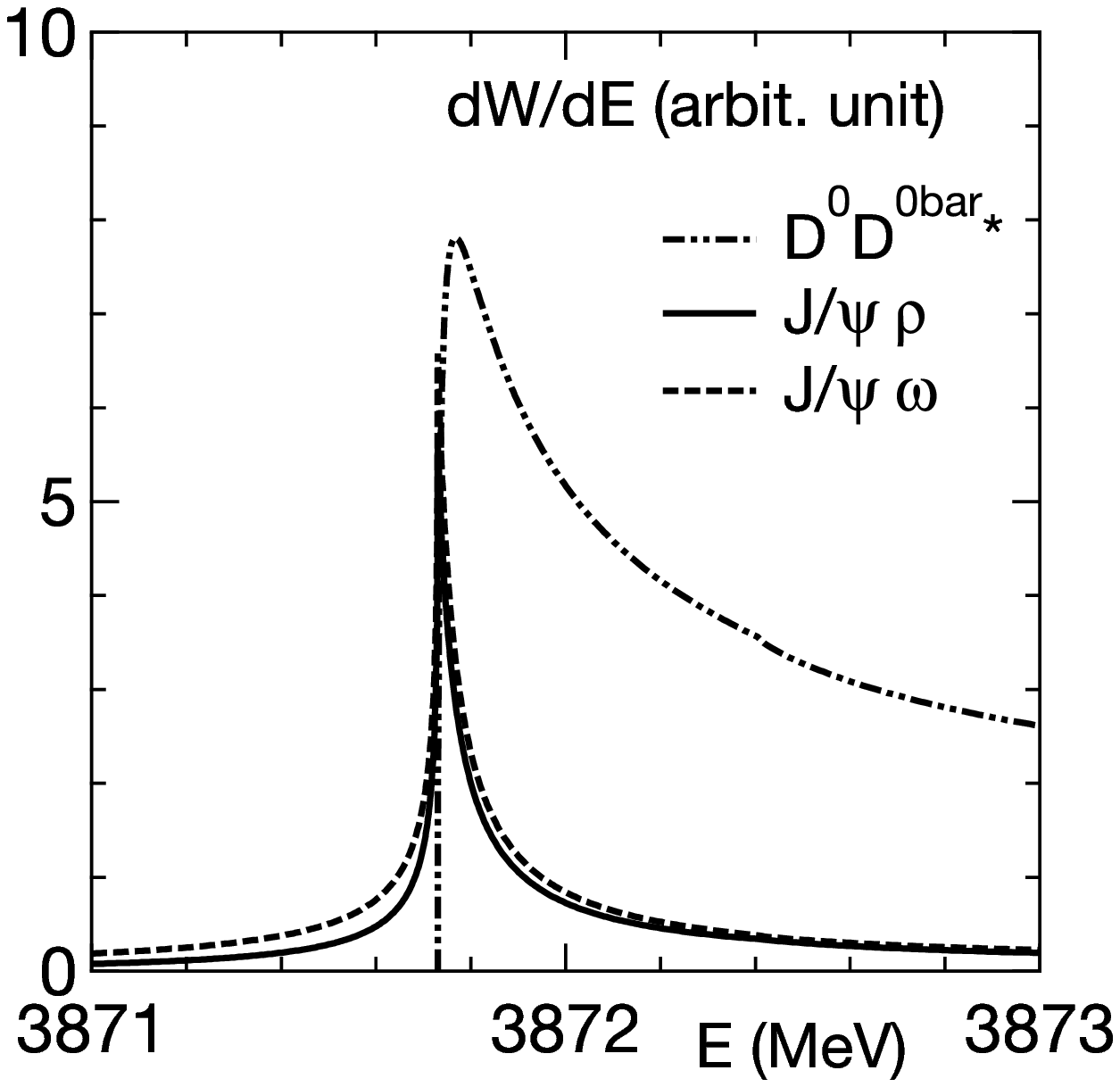}
  \includegraphics[scale=0.4]{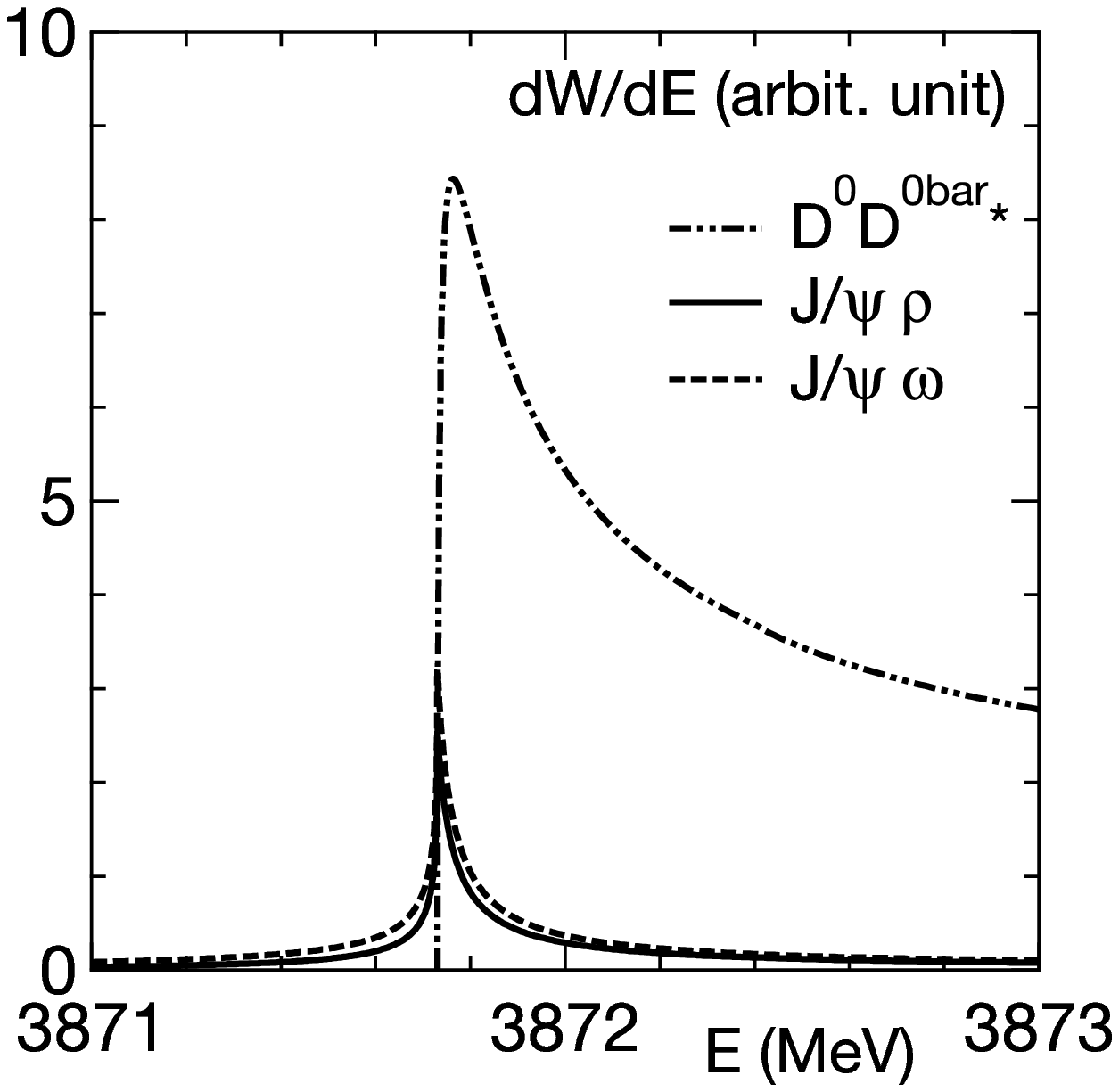}
\caption{The transfer strength from the \ccbar\ core to the
final states for the parameter set (a) A (b) B. }
\label{fig:1}       
\end{figure}

\begin{figure}
\centering
\includegraphics[height=5.2cm,width=5.3cm]{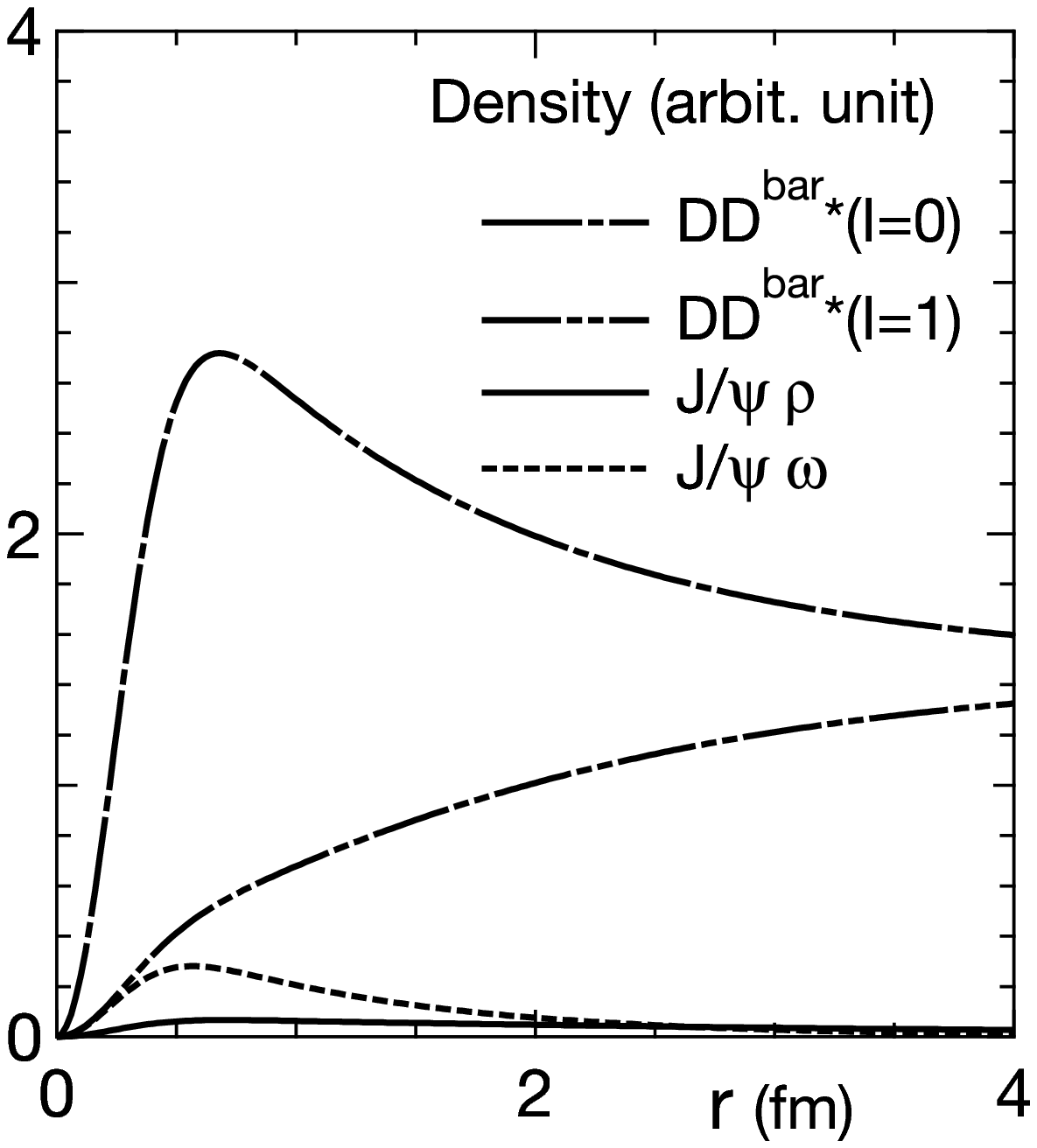}%
\includegraphics[scale=0.37 ]{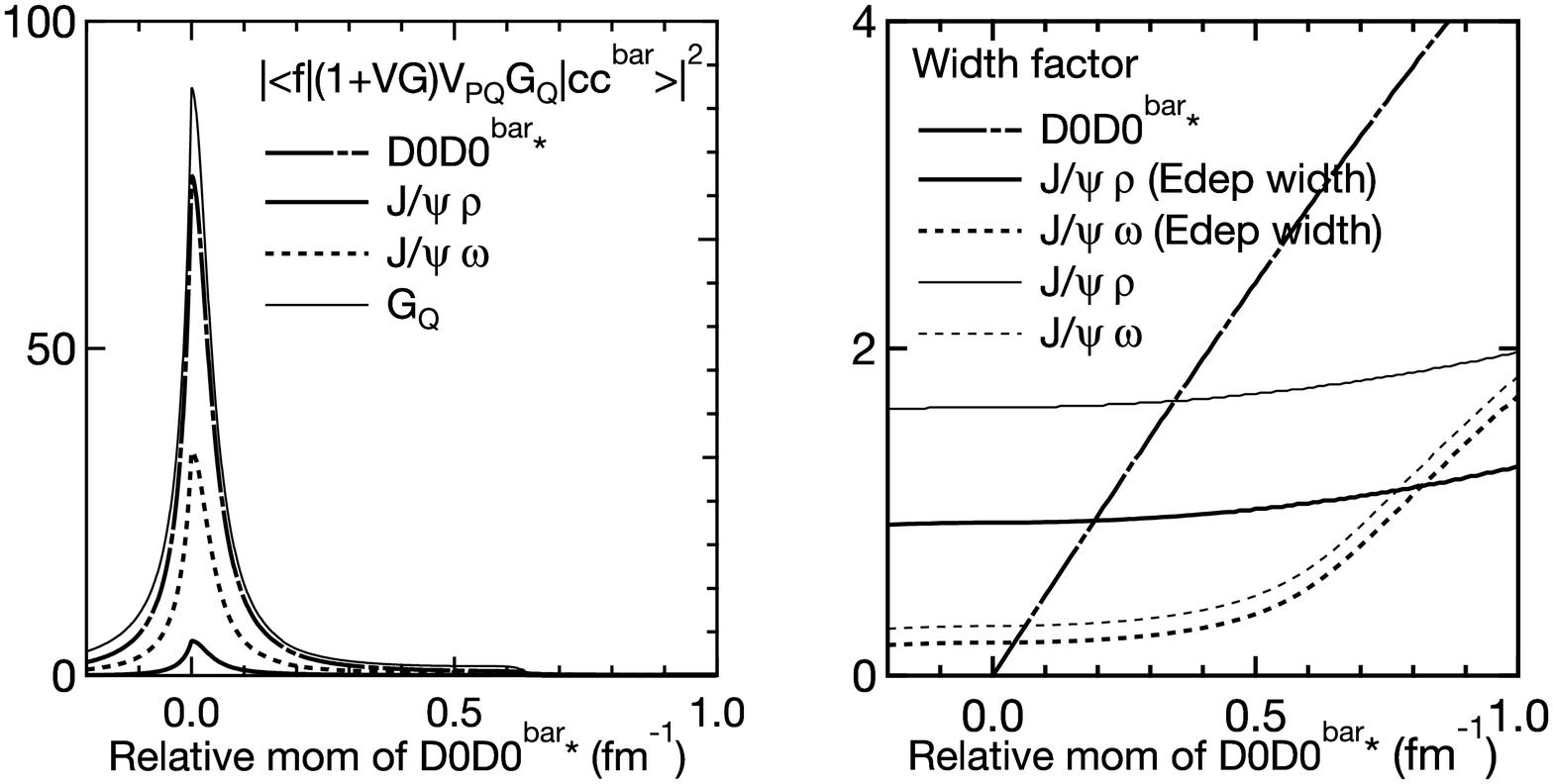}%
\caption{(a) The density of the resonance state,  
(b) the matrix elements for the transition (see text),
and (c) the factors from the meson width.}
\label{fig:2}       
\end{figure}

\section{Results}
In Fig.\ \ref{fig:1}, we show the
transfer strength from the \ccbar\ core to the final two-meson states 
with the parameter set A or B.
In both of the cases, the strengths of $\Jpsi\rho$ and $\Jpsi\omega$
 make a very thin peak at the \DDbarz\ threshold, and their sizes are 
comparable to each other.
Though the model parameters we take are mostly empirical, 
it is found this feature appears for various kinds of model parameters,
provided that the
strength of \DDbarz\ gathers closely above the 
threshold.

In Fig.\ \ref{fig:2}(a), we show the density of the state at the
\DDbarz\ peak energy, 0.05 MeV above the threshold, for the parameter set A.
The isospin 1 component of $D\overline{D}^*$ is
 small at the short distance but becomes the same amount as that of 
the isospin 0 component at $r\rightarrow \infty$, because
only the \DDbarz\ channel is open at this energy.

In Fig.\ \ref{fig:2} (b), we plot  $|\bra \ccbar | G_Q | \ccbar\ket|^2$
and other matrix elements  for the transition from \ccbar\ to the final two-meson channels,
$| \bra f |g^{-1}(1+V_{PP}\GinP)V_{PQ} G_Q | \ccbar\ket|^2$.
In Fig.\ \ref{fig:2}(c), we show the factor from the meson width in the final states
weighted by the interaction range:
$
\int\! { \mu_f\Gamma_V(s(k))
\over
(\kf^2-k^2)^2 +(\mu_f\Gamma_V(s(k)))^2}\,g(k)^2k^2 \rmd k
$.
This factor for \DDbarz\ becomes linear  because it becomes $\kf \,g(\kf)^2$ at $\Gamma_V\rightarrow 0$. As for the vector mesons, the factor is roughly proportional to the width $\Gamma_V$ in the nominator.
It is found 
that the peak feature of \X\ comes from the pole in $G_Q$. 
 The isospin 1 component originates
from the threshold difference in the propagator $\GinP$,
 which is enhanced by the broad width $\Gamma_\rho$;
the factor for the $\rho$ meson is by about five times larger than that of the $\omega$ meson.

\section{Summary}

We solve the Lippmann-Schwinger equation 
for the coupled-channel two-meson scattering problem 
(\DDbarz, \DDbarpm, $\Jpsi\rho$, and $\Jpsi\omega$)
with the \ccbar\  core to investigate the features of \X.
The isospin breaking in the present model comes from  the
difference in the meson masses.
The effects of the vector meson widths
are also taken into account.

It is found that the transfer strength from
the \ccbar\ core to each of
the $\Jpsi\rho$ or $\Jpsi\omega$
has a peak on the \DDbarz\ threshold.
The large width of the $\rho$ meson enhances the isospin 1 component.
The size of the $\Jpsi\rho$ peak is almost the same as that of the 
$\Jpsi\omega$ peak, 
which is consistent with the observed feature.

\begin{acknowledgements}
This work is partly supported by Grants-in-Aid
for scientific research  (20540281 and 21105006).
\end{acknowledgements}



\end{document}